\documentclass[preprint,eqsecnum,aps, showpacs]{revtex4}
\usepackage{graphicx}%
\usepackage{dcolumn}
\usepackage{amsmath}
\usepackage{bm}

\begin{document}

\title{Two neutrino double $\beta $ decay of 94$\leq A\leq $ 110 nuclei for 0$%
^{+}\rightarrow $0$^{+}$ transition}
\author{R. Chandra}
\affiliation{Department of Physics, University of Lucknow, Lucknow-226007, India.\\
}
\author{J. Singh}
\affiliation{Department of Physics, University of Lucknow, Lucknow-226007, India.\\
}
\author{P. K. Rath}
\affiliation{Department of Physics, University of Lucknow, Lucknow-226007, India.\\
}
\author{P. K. Raina}
\affiliation{Department of Physics and Meteorology, IIT Kharagpur-721302,
India.\\}

\author{J. G. Hirsch} 
\affiliation{Instituto de Ciencias Nucleares, Universidad Nacional Aut\'onoma de
M\'exico, A.P. 70-543 M\'exico 04510 D.F.\\ }


\begin{abstract}
The two neutrino double beta decay of $\ ^{94,96}$Zr$,^{98,100}$Mo$,^{104}$%
Ru and $\,^{110}$Pd nuclei for $0^{+}\to 0^{+}$ transition is studied in the
PHFB model in conjunction with the summation method. In the first step, the
reliability of the intrinsic wave functions has been established by
obtaining an overall agreement between a number of theoretically calculated
spectroscopic properties and the available experimental data for $\ ^{94,96}$%
Zr$,^{94,96,98,100}$Mo, $^{98,100,104}$Ru, $^{104,110}$Pd and $^{110}$Cd 
isotopes. Subsequently, the PHFB wave functions of the above
mentioned nuclei are employed to calculate the nuclear transition
matrix elements $M_{2\nu }$ as well as half-lives $T{_{1/2}^{2\nu }}$.
Further, we have studied the effects of deformation on the $M_{2\nu }$.
\end{abstract}

\maketitle 
\section{INTRODUCTION}

\label{sec:level1}The implications of present studies about nuclear double
beta ($\beta \beta $) decay 
\cite{may35, fur39}
are far reaching in nature. The 
two neutrino double beta (2$\nu $ $%
\beta \beta $) decay is a second order process of weak interaction and
conserves the lepton number exactly. Hence, it is allowed in the standard
model of electroweak unification (SM). The half-life of 2$\nu $ $\beta \beta 
$ decay, which is a product of accurately known phase space factor and
appropriate nuclear transition matrix element (NTME) $M_{2\nu }$, has been
already measured for about ten nuclei out of 35 possible candidates. So, the
values of NTME $M_{2\nu }$ can be extracted directly. Consequently, the
validity of different models employed for nuclear structure calculations can
be tested by calculating the $M_{2\nu }$. The 
neutrinoless double beta (0$\nu $ $\beta \beta $) decay
is a convenient tool to test physics beyond the SM. The experimental as well
as theoretical aspect of nuclear $\beta \beta $ decay have been 
widely
reviewed
over the past years [3-19].

Klapdor and his group have recently reported that the 0$\nu $ $\beta \beta $
decay has been observed in $^{76}$Ge. The results are controversial but it
is expected that the issue will be settled soon \cite{kla02}. The aim of all
the present experimental activities is to observe the 0$\nu $ $\beta \beta $
decay. As the 0$\nu $ $\beta \beta $ decay 
has not been observed so far, the
nuclear models predict half-lives assuming certain value for the neutrino
mass or conversely extract various parameters from the observed limits on
half-lives of the 0$\nu $ $\beta \beta $ decay. The reliability of
predictions can be judged \textit{a priori} only from the success of a
nuclear model in explaining various observed physical properties of nuclei.
The common practice is to calculate the $M_{2\nu }$ to start with and
compare with the experimentally observed value as the two decay modes
involve the same set of initial and final nuclear wave functions.

In 2$\nu $ $\beta \beta $ decay, the total angular momentum of four \textit{S%
}-wave leptons can be 0, 1 or 2, and is equal to the total angular momentum
transferred between the parent and daughter nuclei. The lowest 1$^{+}$ state
in the final nucleus of any $\beta \beta $ decay candidate lies much higher
in energy
than the first excited 2$^{+}$ state. Hence, the 0$^{+}\rightarrow $1$^{+}$
transition is much less probable than the 0$^{+}\rightarrow $0$^{+}$ and 0$%
^{+}\rightarrow $2$^{+}$ transitions. Since, the 0$^{+}\rightarrow $2$^{+}$
transition 
has not been detected up to now,
the present theoretical predictions can 
only
be checked against the 0$^{+}\rightarrow $0$^{+}$ transition of 2$\nu $ $\beta \beta $ decay.

In all cases of the $2\nu $ $\beta \beta $ decay for 0$^{+}\rightarrow $0$%
^{+}$ transition, it is observed that the NTMEs $M_{2\nu }$ are 
quenched, 
i.e. they are smaller than those predicted for pure quaisparticle transitions.
 The main objective of all nuclear structure calculations is to
understand the physical mechanism responsible for the suppression of the $%
M_{2\nu }$. Over the past few years, the $M_{2\nu }$ has been calculated
mainly in three types of models, namely the shell model and its variants, the
quasiparticle random phase approximation (QRPA) and its extensions and the
alternative models. In the recent past, the details about these models
-their advantages as well as shortcomings- have been discussed 
by Suhonen \textit{et al. }\cite{suh98} and Faessler \textit{et al.} \cite
{fae98}.

The shell model attempts to solve the nuclear many-body problem as exactly
as possible. Hence, it is the best choice for the calculation of the NTMEs.
However, most of the $\beta \beta $ decay emitters are medium or heavy mass
nuclei for which the number of basis states increases quite drastically. Few
years back, it was not possible to perform a reliable shell-model
calculation beyond the \textit{pf}-shell. Hence, Haxton and Stephenson, Jr. 
\cite{hax84} and Vergados \cite{ver86} have studied the $\beta \beta $\
decay of $^{76}$Ge, $^{82}$Se and $^{128,130}$Te\ nuclei in weak coupling
limit. Recent large scale shell model calculations are more promising in
nature \cite{zha90,cau96}. The calculations by Caurier \textit{et al.} are
more realistic in which the $M_{2\nu }$ of $^{82}$Se is calculated exactly
and those of $^{76}$Ge and $^{136}$Xe are dealt in a nearly exact manner 
\cite{cau96}. The conventional shell model and Monte-Carlo shell model
(MCSM) \cite{rad96} have been tested against each other in case of $^{48}$Ca
and $^{76}$Ge and the agreement is interestingly good. Hence, it is expected
that the MCSM could be a good alternative to conventional shell model
calculations in near future.

Vogel and Zirnbauer were the first to provide an 
explanation of the
observed suppression of $M_{2\nu }$ in the QRPA model by a proper inclusion
of ground state correlations through the 
proton-neutron
\textit{p-p} interaction in the 
\textit{S}=1, \textit{T}=0 channel and the calculated half-lives are in
close agreement with all the experimental data \cite{vog86}. The QRPA
frequently overestimates the ground state correlations as a result of an
increase in the strength of attractive proton-neutron interaction leading to
the collapse of QRPA solutions. The physical value of this force is usually
close to the point at which the QRPA solutions collapse. To cure the strong
suppression of $M_{2\nu }$, several extensions of QRPA have been proposed.
The most important proposals are inclusion of proton-neutron pairing,
renormalized QRPA, higher QRPA, multiple commutator method (MCM) and
particle number projection. However, none of the above methods is free from
ambiguities \cite{fae98}. 
Alternative models, as the operator
expansion method (OEM), the broken SU(4) symmetry, two vacua RPA, the pseudo
SU(3) and the single state dominance hypothesis (SSDH) have their own
problems \cite{suh98}.

The basic 
aim of nuclear many body theory is to 
describe 
as much
observed properties of nuclei
as possible in a coherent 
frame. 
The $\beta \beta $ decay 
can be studied in the same framework of many other nuclear 
properties and decays. 
Over the past years, a vast amount
of data has been collected through experimental studies involving in-beam $%
\gamma $-ray spectroscopy concerning the level energies as well as
electromagnetic properties. The availability of data permits a rigorous and
detailed critique of the ingredients of the microscopic framework that seeks
to provide a description of nuclear $\beta \beta $ decay. However, most of
the calculations of $\beta \beta $ decay matrix elements performed so far do
not satisfy this criterion. Our aim is to study the 2$\nu $ $\beta \beta $
decay of $^{94,96}$Zr, $^{98,100}$Mo, $^{104}$Ru and $^{110}$Pd isotopes for 
$0^{+}\to 0^{+}$ transition not in isolation but together with other
observed nuclear phenomena. The 2$\nu $ $\beta \beta $ decay of $^{100}$Mo
along with the spectroscopic properties has been already studied in the
Projected Hartree-Fock-Bogoliubov (PHFB) model using the closure
approximation \cite{dix02, cha03}. In the present calculation, we have
avoided the closure approximation by making use of the 
summation method \cite{civ93}. Further, the HFB wave function of $^{100}$Mo
are generated with improved accuracy.

The structure of nuclei in the mass region \textit{A}$\approx $100
involving Zr, Mo, Ru, Pd and Cd isotopes is quite complex. With the
discovery of a new region of deformation around \textit{A}= 100 by Cheifetz 
\textit{et al. }\cite{che70}, a well developed rotational spectra was
observed in several neutron rich Mo and Ru isotopes during a study of
fission fragments of $^{252}$Cf. The $B(E2$:$0^{+}\to 2^{+})$ values were
observed to be as enhanced as in the rare-earth and actinide regions. This
mass region offered a nice example of shape transition,
the sudden onset of deformation at neutron number \textit{N}=60. The nuclei are soft
vibrators for neutron number $ N <$ 60 and quasi-rotors for 
$ N >$ 60. The nuclei with neutron number \textit{N}=60 are
transitional nuclei. Thus, in this mass region $^{100}$Zr, $^{102}$Mo, $%
^{104}$Ru and $^{106}$Pd are observed to be transitional cases. In case of
Cd isotopes, similar shape transition occurs at \textit{A}=100. Hence, it is
expected that deformation will play a crucial role in reproducing the
properties of nuclei in this mass region \textit{A}$\approx $100.
Moreover, it has been already conjectured that the deformation can play a
crucial role in case of $\beta \beta $ decay of $^{100}$Mo and $^{150}$Nd 
\cite{gri92,suh94}. Further, all the nuclei undergoing $\beta \beta $ decay
are even-even type, in which the pairing degrees of freedom play an
important role. Hence, it is desirable to have a model which incorporates
the pairing and deformation degrees of freedom on equal footing in its
formalism. For this purpose, the PHFB model is one of the most natural
choices.

Over the past twenty years, extensive studies of shape transition vis-a-vis
electromagnetic properties of Zr and Mo isotopes have been successfully
carried out in the PHFB model \cite{kho82} using the pairing plus
quadrupole-quadrupole (PPQQ) interaction \cite{bar68}. The success of the
PHFB model in explaining the observed experimental trends in the mass region 
\textit{A}$\approx $100 motivated us to apply the HFB wave functions to
study the nuclear 2$\nu $ $\beta \beta $ decay of $^{100}$Mo$\to $ $^{100}$%
Ru for 0$^{+}\rightarrow $0$^{+}$ transition. Further, the success of the
PHFB model in conjunction with the PPQQ interaction in explaining the yrast
spectra, reduced transition probabilities $B(E2$:$0^{+}\rightarrow 2^{+})$,
static quadrupole moments $Q(2^{+})$, $g$-factors $g(2^{+})$ of $^{100}$Mo
and $^{100}$Ru nuclei as well as the $T_{1/2}^{2\nu }$(0$^{+}\rightarrow $0$%
^{+}$) of $^{100}$Mo \cite{dix02} has prompted us to apply the PHFB model to
study the 2$\nu $ $\beta \beta $ decay of some nuclei namely $^{94,96}$Zr, $%
^{98,100}$Mo, $^{104}$Ru and $^{110}$Pd for 0$^{+}\rightarrow $0$^{+}$
transition in the mass range 94$\leq $\textit{A}$\leq $110.

It is well known that the pairing part of the interaction (\textit{P})
accounts for the sphericity of nucleus, whereas the quadrupole-quadrupole (%
\textit{QQ}) interaction increases the collectivity in the nuclear intrinsic
wave functions and makes the nucleus deformed. Hence, the PHFB model using
the PPQQ interaction is a convenient choice to examine the explicit role of
deformation on NTMEs $M_{2\nu }.$ In case of $^{100}$Mo for 0$%
^{+}\rightarrow $0$^{+}$ transition, we have observed that the deformation
plays an important role in reproducing a realistic $M_{2\nu }$ \cite{dix02}.
Therefore, we have also studied the variation of $M_{2\nu }$ vis-a-vis the
change in deformation through the changing strength of the \textit{QQ}
interaction.

The present paper has been organized as follows. The theoretical formalism
to calculate the half-life of 2$\nu $ $\beta \beta $ decay mode has been
given by Haxton and Stephenson, Jr. \cite{hax84}, Doi \textit{et al. }\cite
{doi85} and Tomoda \cite{tom91}. Hence in Sec. II, we briefly outline steps
of the above derivations for clarity in notations used in the present paper
following Doi \textit{et al. }\cite{doi85}. Further, we have presented
formulae to calculate the NTME of the 2$\nu $ $\beta \beta $\ decay in the
PHFB model in conjunction with the summation method. Expressions used to
calculate the nuclear spectroscopic properties, namely yrast spectra,
reduced $B(E2)$ transition probabilities, static quadrupole moments and $g$%
-factors have been given by Dixit \textit{et al. }\cite{dix02}. In Sec. III
A, as a test of the reliability of the wave functions, we have calculated
the yrast spectra, reduced $B(E2$:$0^{+}\rightarrow 2^{+})$ transition
probabilities, static quadrupole moments $Q(2^{+})$ and $g$-factors $%
g(2^{+}) $ of nuclei participating in the 2$\nu $ $\beta \beta $ decay and
compared with the available experimental data. Subsequently, the HFB wave
functions of the above mentioned nuclei\textrm{\ }are employed to calculate
the $M_{2\nu }$ as well as half-lives $T{_{1/2}^{2\nu }}$ in Sec. III B. In
Sec. III C, the role of deformation on $M_{2\nu }$ have been studied through
varying strength of the \textit{QQ} interaction. Finally, the conclusions
are given in Sec. IV.

\section{THEORETICAL FRAMEWORK}

The inverse half-life of the $\ $2$\nu $ $\beta \beta $\ decay for $\ $0$%
^{+}\to $0$^{+}$\ transition is given by 
\begin{equation}
\lbrack T_{1/2}^{2\nu }(0^{+}\to 0^{+})]^{-1}=G_{2\nu }|M_{2\nu }|^{2}
\end{equation}
The integrated kinematical factor $\ G_{2\nu }$\ can be calculated with good
accuracy \cite{doi85} and the NTME $M_{2\nu }$ is given by 
\begin{eqnarray}
M_{2\nu } &=&\sum\limits_{N}\frac{\langle 0_{F}^{+}||\bm{\sigma }\tau
^{+}||1_{N}^{+}\rangle \langle 1_{N}^{+}||\bm{\sigma }\tau
^{+}||0_{I}^{+}\rangle }{E_{N}-(E_{I}+E_{F})/2} \\
&=&\sum\limits_{N}\frac{\langle 0_{F}^{+}||\bm{\sigma }\tau
^{+}||1_{N}^{+}\rangle \langle 1_{N}^{+}||\bm{\sigma }\tau
^{+}||0_{I}^{+}\rangle }{E_{0}+E_{N}-E_{I}}
\end{eqnarray}
where $\ $%
\begin{equation}
E_{0}=\frac{1}{2}\left( E_{I}-E_{F}\right) =\frac{1}{2}Q_{\beta \beta }+m_{e}
\end{equation}

The summation over intermediate states can be completed using the summation
method \cite{civ93} and the $M_{2\nu }$ can be written as 
\begin{equation}
M_{2\nu }=\frac{1}{E_{0}}\left\langle 0_{F}^{+}\left| \sum_{m}(-1)^{m}\Gamma
_{-m}F_{m}\right| 0_{I}^{+}\right\rangle 
\end{equation}
where the Gamow-Teller (GT) operator $\Gamma _{m}$ is given by 
\begin{equation}
\Gamma _{m}=\sum_{s}\sigma _{ms}\tau _{s}^{+}
\end{equation}
and 
\begin{equation}
F_{m}=\sum_{\lambda =0}^{\infty }\frac{(-1)^{\lambda }}{E_{0}^{\lambda }}%
D_{\lambda }\Gamma _{m}
\end{equation}
with 
\begin{equation}
D_{\lambda }\Gamma _{m}=\left[ H,\left[ H,........,\right. \left[ H,\Gamma
_{m}\right] .......\right] ^{(\lambda \hbox{ times})}  \label{eqcom}
\end{equation}
Presently, we have used a Hamiltonian with PPQQ type \cite{bar68} of
effective two-body interaction, which is explicitly written as 
\begin{equation}
{H}=H_{sp}+V(P)+\chi _{qq}V(QQ)
\end{equation}
where $H_{sp}$ denotes the single particle Hamiltonian. The pairing part of
the effective two-body interaction $V(P)$ is written as 
\begin{equation}
V{(}P{)}=-\left( \frac{G}{4}\right) \sum\limits_{\alpha \beta
}(-1)^{j_{\alpha }+j_{\beta }-m_{\alpha }-m_{\beta }}a_{\alpha }^{\dagger
}a_{\bar{\alpha}}^{\dagger }a_{\bar{\beta}}a_{\beta }
\end{equation}
where $\alpha $ denotes the quantum numbers (\textit{nljm}). The state $\bar{%
\alpha}$ is same as $\alpha $ but with the sign of \textit{m} reversed. The 
\textit{QQ} part of the effective interaction $V(QQ)$\ is given by 
\begin{equation}
V(QQ)=-\left( \frac{\chi }{2}\right) \sum\limits_{\alpha \beta \gamma \delta
}\sum\limits_{\mu }(-1)^{\mu }\langle \alpha |q_{\mu }^{2}|\gamma \rangle
\langle \beta |q_{-\mu }^{2}|\delta \rangle \ a_{\alpha }^{\dagger }a_{\beta
}^{\dagger }\ a_{\delta }\ a_{\gamma }
\end{equation}
where 
\begin{equation}
{q_{\mu }^{2}}=\left( \frac{16\pi }{5}\right) ^{1/2}r^{2}Y_{\mu }^{2}(\theta
,\phi )
\end{equation}
The $\ \chi _{qq}$ is an arbitrary parameter and the final results are
obtained by setting the $\ \chi _{qq}$ = 1. The purpose of introducing $\chi
_{qq}$ is to study the role of deformation by varying the strength of 
\textit{QQ} interaction.

When the GT operator commutes with the effective two-body interaction,
the Eq. (\ref{eqcom}) can be further simplified to 
\begin{equation}
M_{2\nu }=\sum\limits_{\pi ,\nu }\frac{\langle 0_{F}^{+}||\bm{\sigma
.\sigma }\tau ^{+}\tau ^{+}||0_{I}^{+}\rangle }{E_{0}+\varepsilon (n_{\pi
},l_{\pi },j_{\pi })-\varepsilon (n_{\nu },l_{\nu },j_{\nu })}
\end{equation}
In
the case of the pseudo SU(3) model [33-35], the GT operator commutes with the
two-body interaction and the energy denominator is a well-defined quantity
without any free parameter. It has been evaluated exactly for 2$\nu $ $\beta
^{-}\beta ^{-}$ \cite{cas94,hir95a} and 2$\nu $ ECEC modes 
\cite{cer99} in the context of pseudo SU(3) scheme. In the present case, the
model Hamiltonian is not isospin symmetric. Hence, the energy denominator 
has not the simple form shown in Eq. (2.13).
However, the violation of isospin symmetry for
the \textit{QQ } part of our model Hamiltonian is negligible as will be evident from
the parameters of the two-body interaction given later. 
Also the violation
of isospin symmetry for the pairing part of the two-body interaction is
presumably small. With these assumptions, the expression to calculate the
NTME $M_{2\nu }$ of 2$\nu $ $\beta \beta $ decay for 0$^{+}\rightarrow 0^{+}$
transition in the PHFB model is obtained as follows.

The axially symmetric HFB intrinsic state with \textit{K}=0 can be written
as 
\begin{equation}
{|\Phi _{0}\rangle }=\Pi _{im}(u_{im}+v_{im}b_{im}^{\dagger }b_{i\bar{m}%
}^{\dagger })|0\rangle
\end{equation}
where the creation operators $\ b_{im}^{\dagger }$\ and $\ b_{i\bar{m}%
}^{\dagger }$\ are given by 
\begin{equation}
b_{im}^{\dagger }=\sum\limits_{\alpha }C_{i\alpha ,m}a_{\alpha m}^{\dagger }%
\hbox{ and }b_{i\bar{m}}^{\dagger }=\sum\limits_{\alpha
}(-1)^{l+j-m}C_{i\alpha ,m}a_{\alpha ,-m}^{\dagger }
\end{equation}
Using the standard projection technique, a state with good angular momentum $%
\mathbf{J}$ is obtained from the HFB intrinsic state through the following
relation.

\begin{eqnarray}
|\Psi _{MK}^{J}\rangle &=&P_{MK}^{J}|\Phi _{K}\rangle  \nonumber \\
&=&\left[ \frac{(2J+1)}{{8\pi ^{2}}}\right] \int D_{MK}^{J}(\Omega )R(\Omega
)|\Phi _{K}\rangle d\Omega
\end{eqnarray}
where $\ R(\Omega )$\ and $\ D_{MK}^{J}(\Omega )$\ are the rotation operator
and the rotation matrix respectively.

Finally, one obtains the following expression for the NTME $M_{2\nu }$ of 2$%
\nu $ $\beta \beta $ decay for 0$^{+}\rightarrow 0^{+}$ transition in the
PHFB model using the summation method.

\begin{eqnarray}
M_{2\nu } &=&\sum\limits_{\pi ,\nu }\frac{\langle {\Psi _{00}^{J_{f}=0}}||%
\bm{\sigma .\sigma }\tau ^{+}\tau ^{+}||{\Psi _{00}^{J_{i}=0}}\rangle }{%
E_{0}+\varepsilon (n_{\pi },l_{\pi },j_{\pi })-\varepsilon (n_{\nu },l_{\nu
},j_{\nu })}  \nonumber \\
&=&\left[ n_{(Z,N)}^{Ji=0}n_{(Z+2,N-2)}^{J_{f}=0}\right]
^{-1/2}\int\limits_{0}^{\pi }n_{(Z,N),(Z+2,N-2)}(\theta )  \nonumber \\
&&\times \sum\limits_{\alpha \beta \gamma \delta }\frac{\left\langle \alpha
\beta \left| \bm{\sigma }_{1}.\bm{\sigma }_{2}\tau ^{+}\tau
^{+}\right| \gamma \delta \right\rangle }{E_{0}+\varepsilon _{\alpha
}(n_{\pi },l_{\pi },j_{\pi })-\varepsilon _{\gamma }(n_{\nu },l_{\nu
},j_{\nu })}  \nonumber \\
&&\times \sum\limits_{\varepsilon \eta }\left[ \left( 1+F_{Z,N}^{(\pi
)}(\theta )f_{Z+2,N-2}^{(\pi )*}\right) \right] _{\varepsilon \alpha
}^{-1}\left( f_{Z+2,N-2}^{(\pi )*}\right) _{\varepsilon \beta }  \nonumber \\
&&\times \left[ \left( 1+F_{Z,N}^{(\nu )}(\theta )f_{Z+2,N-2}^{(\nu
)*}\right) \right] _{\gamma \eta }^{-1}\left( F_{Z,N}^{(\nu )*}\right)
_{\eta \delta }sin\theta d\theta
\end{eqnarray}
where 
\[
n^{J}=\int\limits_{0}^{\pi }\left[ det\left( 1+F^{(\pi )}f^{(\pi )^{\dagger
}}\right) \right] ^{1/2}\left[ det\left( 1+F^{(\nu )}f^{(\nu )^{\dagger
}}\right) \right] ^{1/2}d_{00}^{J}(\theta )sin(\theta )d\theta 
\]
and 
\begin{equation}
n_{(Z,N),(Z+2,N-2)}{(\theta )}=\left[ det\left( 1+F_{Z,N}^{(\nu
)}f_{Z+2,N-2}^{(\nu )^{\dagger }}\right) \right] ^{1/2}\times \left[
det\left( 1+F_{Z,N}^{(\pi )}f_{Z+2,N-2}^{(\pi )^{\dagger }}\right) \right]
^{1/2}
\end{equation}
The $\pi (\nu )$\ represents the proton (neutron) of nuclei involved in the 2%
$\nu $ $\beta \beta $ decay process. The matrices $F_{Z,N}(\theta )$\ and $\
f_{Z,N}$\ are given by 
\begin{equation}
F_{Z,N}(\theta )=\sum\limits_{m_{\alpha }^{\prime }m_{\beta }^{\prime
}}d_{m_{\alpha },m_{\alpha }^{\prime }}^{j_{\alpha }}(\theta )d_{m_{\beta
},m_{\beta }^{\prime }}^{j_{\beta }}(\theta )f_{j_{\alpha }m_{\alpha
}^{\prime },j_{\beta }m_{\beta }^{\prime }}
\end{equation}
\begin{equation}
f_{Z,N}=\sum\limits_{i}C_{ij_{\alpha },m_{\alpha }}C_{ij_{\beta },m_{\beta
}}\left( v_{im_{\alpha }}/u_{im_{\alpha }}\right) \delta _{m_{\alpha
},-m_{\beta }}
\end{equation}
The required NTME $M_{2\nu }$ is calculated using the results of PHFB
calculations which are summarized by the amplitudes $(u_{im},v_{im})$\ and
the expansion coefficients $C_{ij,m}$. In the first step, matrices $\ F^{\pi
,\nu }$ and $\ f^{\pi ,\nu }$ are setup for the nuclei involved in the 2$\nu 
$ $\beta \beta $ decay making use of 20 Gaussian quadrature points in the
range (0, ~$\pi $). Finally using the Eq. (2.17), the required NTME can be
calculated in a straightforward manner.

\section{RESULTS AND DISCUSSIONS}

The model space, single particle energies (SPE's) and two-body interactions
are same as
in 
our earlier calculation on 2$\nu $ $\beta \beta $ decay of $%
^{100}$Mo for 0$^{+}\rightarrow $0$^{+}$ transition \cite{dix02}. However,
we have 
included a brief discussion of them in the following for convenience.
We have treated the doubly even nucleus $^{76}$Sr (\textit{N}=\textit{Z}=38)
as an inert core with the valence space spanned by orbits 1\textit{p}$%
_{1/2}, $ 2\textit{s}$_{1/2,}$ 1\textit{d}$_{3/2}$, 1\textit{d}$_{5/2}$, 0%
\textit{g}$_{7/2}$, 0\textit{g}$_{9/2}$ and 0\textit{h}$_{11/2}$ for protons
and neutrons. The orbit 1\textit{p}$_{1/2}$ has been included in the valence
space to examine the role of the \textit{Z}=40 proton core vis-a-vis the
onset of deformation in the highly neutron rich isotopes.

The set of single particle energies (SPE's) used here are (in MeV) $%
\varepsilon $(1\textit{p}$_{1/2}$)=-0.8, $\varepsilon $(0\textit{g}$_{9/2}$%
)=0.0, $\varepsilon $(1\textit{d}$_{5/2}$)=5.4, $\varepsilon $(2\textit{s}$%
_{1/2}$)=6.4, $\varepsilon $(1\textit{d}$_{3/2}$)=7.9, $\varepsilon $(0%
\textit{g}$_{7/2}$)=8.4 and $\varepsilon $(0\textit{h}$_{11/2}$)=8.6 for
protons and neutrons. This set of SPE's but for the $\varepsilon $(0\textit{h%
}$_{11/2}$), which is slightly lowered, has been employed in a number of
successful shell model \cite{ver71} as well as variational model
calculations \cite{kho82} for nuclear properties in the mass region \textit{A%
}=100. The strengths of the pairing interaction is fixed through the
relation $G_{p}$ =30/\textit{A} MeV and $G_{n}$=20/\textit{A} MeV, which are
same as used by Heestand \textit{et al. }\cite{hee69} to explain the
experimental $g(2^{+})$ data of some even-even Ge, Se, Mo, Ru, Pd, Cd and Te
isotopes in Greiner's collective model \cite{gre66}. For $^{96}$Zr, we have
used $G_{n}$=22/\textit{A} MeV. The strengths of the like particle
components of the \textit{QQ} interaction are taken as: $\chi _{pp}$ = $\chi
_{nn}$ = 0.0105 MeV \textit{b}$^{-4}$, where \textit{b} is oscillator
parameter.

The strength of proton-neutron (\textit{pn}) component of the \textit{QQ}
interaction $\chi _{pn}$ is varied so as to obtain the spectra of considered
nuclei namely $^{94,96}$Zr, $^{94,96,98,100}$Mo, $^{98,100,104}$Ru, $%
^{104,110}$Pd and $^{110}$Cd in optimum agreement with the experimental
results. To be more specific, we have taken the theoretical spectra to be
the optimum one if the excitation energy of the $\ $2$^{+}$ state \ $%
E_{2^{+}}$ is reproduced as closely as possible to the experimental value.
Thus for a given model space, SPE's, $G_{p}$, $G_{n}$ and $\chi _{pp}$, we
have fixed $\chi _{pn}$ through the experimentally available energy spectra.
We have given the values of $\chi _{pn}$ in Table I. These values for the
strength of the \textit{QQ} interaction are comparable to those suggested by
Arima on the basis of an empirical analysis of the effective two-body
interactions \cite{ari81}. All the parameters are kept fixed throughout the
calculation.

\subsection{The yrast spectra and electromagnetic properties}

In Table I, we have presented yrast energies for $\ E_{2^{+}}$ to $\
E_{6^{+}}$ levels of all nuclei of interest. 
The agreement between the theoretically reproduced \ $E_{2^{+}}$ and the
experimentally observed \ $E_{2^{+}}$ \cite{sak84} is quite good. 
However, 
it is observed that in comparison to the experimental spectra, the
theoretical spectra is more expanded. This can be
corrected to some extent
in the PHFB model in conjunction with the VAP prescription \cite{kho82}.
However, our aim is to reproduce properties of the low-lying 2$^{+}$ state.
Hence, we have not attempted to invoke the VAP prescription, which will
unnecessarily complicate the calculations.

In Table II we have presented the calculated as well as the experimentally
observed values of the reduced transition probabilities $\ B(E2$:$0^{+}\to
2^{+})$ \cite{ram87}, static quadrupole moments $\ Q(2^{+})$ and the
gyromagnetic factors $\ g(2^{+})$ \cite{rag89}. We have given $B(E2$:$%
0^{+}\to 2^{+})$ results for effective charges $\ e_{eff}$ =0.40, 0.50 and
0.60 in columns 2 to 4 respectively. The experimentally observed values are
displayed in column 5. In case of $B(E2$:$0^{+}\to 2^{+})$, only some
experimentally observed representative values are tabulated. It is noticed
that the calculated values are in excellent agreement with the observed $%
B(E2 $:$0^{+}\to 2^{+})$ in case of $^{94}$Zr, $^{94,100}$Mo, $^{100,104}$Ru
and $^{104}$Pd nuclei for $\ e_{eff}$ =0.60. The calculated and observed $%
B(E2$:$0^{+}\to 2^{+})$ values are again in agreement in case of $^{96}$Zr
and $^{96}$Mo nuclei for $\ e_{eff}$ =0.50. The calculated $B(E2$:$0^{+}\to
2^{+}) $ values for $e_{eff}$ =0.50 differ by 0.046 and 0.004 $e^{2}$ b$%
^{2}\ $only in case of $^{110}$Pd and $^{110}$Cd nuclei respectively from
the experimental limits. The agreement between the theoretical and
experimental $B(E2$:$0^{+}\to 2^{+})$ values is quite good in case of $^{98}$%
Mo and $^{98}$Ru nuclei for $e_{eff}$ =0.40.

The theoretically calculated $\ Q(2^{+})$ are tabulated in columns 6 to 8
for the same effective charges as given above. The experimental $\ Q(2^{+})$
results are given in column 9. No experimental $Q(2^{+})$ result is
available for $^{94,96}$Zr. It can be seen that for the same effective
charge as used in case of $B(E2$:$0^{+}\to 2^{+})$, the agreement between
the calculated and experimental $Q(2^{+})$ values is quite good for $^{104}$%
Ru and $^{110}$Pd nuclei. The discrepancy between the calculated and
experimental values are off by 0.089, 0.14 and 0.023 e b in case of $%
^{98,100}$Mo and $^{100}$Ru nuclei respectively. The theoretical $Q(2^{+})$
results are quite off from the observed values for the rest of nuclei.

The \ $g(2^{+})$ values are calculated with $\ g_{l}^{\pi }$=1.0, $\
g_{l}^{\nu }$=0.0, $\ g_{s}^{\pi }$=$\ g_{s}^{\nu }$=0.60. No experimental
result is available for $^{96}$Zr and $^{94,96}$Mo. The calculated and
experimentally observed$\ g(2^{+})$ are in good agreement for $^{98,100}$Mo, 
$^{98}$Ru, $^{104}$Pd and $^{110}$Cd nuclei. The discrepancy between the
theoretically calculated and experimentally observed $\ g(2^{+})$ values are
0.035, 0.021 and 0.078 nm only for $^{100,104}$Ru and $^{110}$Pd nuclei
respectively. The theoretical $g(2^{+})$ value of $^{94}$Zr is a
pathological case. The calculated $g(2^{+})$ value is 0.121 nm while the
most recent measured value is -0.329$\pm $0.015 nm \cite{spe02}.

>From the overall agreement between the calculated and observed
electromagnetic properties, it is clear that the PHFB wave functions of $%
^{94,96}$Zr, $^{94,96,98,100}$Mo, $^{98,100,104}$Ru, $^{104,110}$Pd and $%
^{110}$Cd nuclei generated by fixing $\chi _{pn}$ to reproduce the yrast
spectra are quite reliable. Hence, we proceed to calculate the NTMEs $%
M_{2\nu }$ as well as half-lives $T_{1/2}^{2\nu }$ of $^{94,96}$Zr, $%
^{98,100}$Mo, $^{104}$Ru and $^{110}$Pd nuclei for $\ $0$^{+}\to $0$^{+}$
transition.

\subsection{Results of $\ 2\nu $ $\beta \beta $ decay}

The phase space factors $G_{2\nu }$ for 0$^{+}\rightarrow $0$^{+}$
transition have been given by Boehm \textit{et al.} for $\ g_{A}$= 1.25 \cite
{boe92}. These $G_{2\nu }$ are 2.304$\times $10$^{-21}$, 1.927$\times $10$%
^{-17}$, 9.709$\times $10$^{-29}$, 9.434$\times $10$^{-18}$, 9.174$\times $10%
$^{-21}$ and 3.984$\times $10$^{-19}$ yr$^{-1}$ for $^{94,96}$Zr, $^{98,100}$%
Mo, $^{104}$Ru and $^{110}$Pd nuclei respectively. However, in heavy nuclei
it is more justified to use the nuclear matter value of $\ g_{A}$ around
1.0. Hence, the experimental $M_{2\nu }$ as well as the theoretical $%
T_{1/2}^{2\nu }$ are calculated for $\ g_{A}$=1.0 and 1.25.

In Table III, we have compiled all the available experimental and the
theoretical results along with our calculated $M_{2\nu }$ and corresponding
half-lives $\ T_{1/2}^{2\nu }$ of $^{94,96}$Zr, $^{98,100}$Mo, $^{104}$Ru
and $^{110}$Pd isotopes for $\ $0$^{+}\to $0$^{+}$ transition. We have also
presented the $M_{2\nu }$ extracted from the experimentally observed $%
T_{1/2}^{2\nu }$ in Column 5 of Table III using the given phase space
factors. We have presented only the theoretical $T_{1/2}^{2\nu }$ for those
models for which no direct or indirect information about $M_{2\nu }$ is
available to us.

The 2$\nu $ $\beta \beta $ decay of $^{94}$Zr$\to ^{94}$Mo for $\ $0$^{+}\to 
$0$^{+}$ transition has been 
investigated experimentally only by Arnold \cite {arn99}, who reported the limit
$T_{1/2}^{2\nu } > 1.1 \times $10$%
^{17}$ yr. 
Theoretical calculations have been done by employing QRPA 
\cite{sta90}, OEM \cite{hir94}, and SRQRPA \cite{bob00}. The presently
calculated half-life in PHFB model for $g_{A}$=1.25 is 7.51$\times $10$^{22}$
yr, which is closer to the value obtained in QRPA model of Staudt \textit{et
al. }\cite{sta90} and approximately twice of the lower limit given by Bobyk 
\textit{et al. }\cite{bob00}. On the other hand, the calculated half-life $%
T_{1/2}^{2\nu }$ in OEM by Hirsch \textit{et al. }\cite{hir94} is larger
than our PHFB model value for $g_{A}$=1.25 by a factor of 22 approximately.
The predicted $T_{1/2}^{2\nu }$ in PHFB model for $g_{A}$=1.0 is 1.834$%
\times $10$^{23}$ yr.

In case of $^{96}$Zr, all the available experimental [18,44,48-51] and
theoretical results [45-47,52-56] along with our calculated $M_{2\nu }$ and
corresponding $T_{1/2}^{2\nu }$ are compiled in Table III. In comparison to
the experimental $M_{2\nu }$, the theoretically calculated value given by
Stoica using SRPA(WS) \cite{sto95} is too small. On the other hand, the
calculated half-life $T_{1/2}^{2\nu }$ in OEM \cite{hir94} is quite off from
the observed experimental value. The $M_{2\nu }$ calculated by Engel \textit{%
et al.} using QRPA \cite{eng88} and Barabash \textit{et al. }\cite{bar96}
using QRPA (AWS) for $g_{A}$=1.0 is close to the experimentally observed
lower limit of Wieser \textit{et al. }\cite{wie01}. The $T_{1/2}^{2\nu }$
calculated by Toivanen \textit{et al.} in RQRPA (WS) and RQRPA (AWS) are 4.2$%
\times $10$^{19}$ yr and 4.4$\times $10$^{19}$ yr \cite{toi97} respectively
and they are quite close to the experimental value of Kawashima \textit{et
al. }\cite{kaw93}. The predicted half-life $T_{1/2}^{2\nu }$ of Bobyk 
\textit{et al. }\cite{bob00} has a wide range and favor all the available
experimental results. On the other hand, the $T_{1/2}^{2\nu }$ predicted by
Staudt \textit{et al. }\cite{sta90} is in agreement with the experimental
result of Barabash \cite{bar98} and Wieser \textit{et al. }\cite{wie01}.
However, the $T_{1/2}^{2\nu }$ calculated in PHFB model and in SU(4)$%
_{\sigma \tau }$ by Rumyantsev \textit{et al. }\cite{rum98} favor the
experimental values of NEMO \cite{arn99,bar98} and Wieser \textit{et al. }%
\cite{wie01} for g$_{A}$=1.25.

In case of $^{98}$Mo$\to ^{98}$Ru, no experimental result for $T_{1/2}^{2\nu
}$ is available so far. The theoretical calculations have been carried out
in QRPA \cite{sta90}, OEM \cite{hir94} and SRQRPA \cite{bob00}. The
calculated $T_{1/2}^{2\nu }$ for $g_{A}$=1.25 in PHFB model is in the range
given by Bobyk \textit{et al. }in SRQRPA \textit{\ }model\textit{\ }\cite
{bob00}. In the PHFB model for $g_{A}$=1.0, the predicted half-life of 2$\nu 
$ $\beta \beta $ decay $T_{1/2}^{2\nu }$ is 1.49$\times $10$^{30}$ yr. The
predicted $T_{1/2}^{2\nu }$ in QRPA by Staudt \textit{et al. }\cite{sta90}
and in OEM by Hirsch \textit{et al. }\cite{hir94} are larger than our
predicted value for $g_{A}$=1.0 by approximately a factor of 2 and 4
respectively.

The 2$\nu $ $\beta \beta $ decay of $^{100}$Mo for $\ $0$^{+}\to $0$^{+}$
transition have been investigated by many experimental groups [18,51,57-65]
as well as theoreticians by employing different theoretical frameworks
[29,30,45-47,52,53,56,66-68]. In comparison to the experimental $M_{2\nu }$,
the theoretically calculated value given by Stoica using SRPA(WS) \cite
{sto95} is too small. The $\ $2$\nu $ $\beta \beta $ decay rate of $^{100}$%
Mo calculated by Staudt \textit{et al. }\cite{sta90} and Hirsch \textit{et
al.} using OEM \cite{hir94} are off from the experimental $T_{1/2}^{2\nu }$.
For $\ g_{A}$=1.0, the $M_{2\nu }$ calculated by Griffith \textit{et al. }%
\cite{gri92} using QRPA model favors the results of INS Baksan \cite{vas90}
and LBL \cite{gar93} due to large error bar in the experimental $%
T_{1/2}^{2\nu }$. On the other hand, the $M_{2\nu }$ predicted by Engel 
\textit{et al. }\cite{eng88} and Civitarese \textit{et al. }\cite{civ98} for 
$\ g_{A}$=1.0 are in agreement with the results of LBL \cite{gar93}, LBL
collaboration \cite{als97}, UC Irvine \cite{sil97} and ITEP+INFN \cite{ash01}
due to experimental error bars. The values of $M_{2\nu }$ predicted in SU(4)$%
_{\sigma \tau }$ \cite{rum98} and SU3(SPH) \cite{hir95b} are nearly
identical and close to the experimental result given by Vasilev \textit{et
al. }\cite{vas90} and ITEP+INFN \cite{ash01} for $\ g_{A}$=1.25. The same
two $M_{2\nu }$ for $\ g_{A}$=1.0 are in agreement with the results of
UC-Irvine \cite{ell91}, ELEGANTS V, LBL and NEMO. Further, the value of $%
M_{2\nu }$ given by PHFB model, Suhonen \textit{et al.} using QRPA(EMP) \cite
{suh94} and Hirsch \textit{et al.} using SU3(DEF) \cite{hir95b} favor the
results of UC-Irvine (results of Elliott \textit{et al.}) \cite{ell91},
ELEGANTS V \cite{eji91}, LBL \cite{gar93}, NEMO \cite{das95}, LBL
collaboration \cite{als97} and ITEP+INFN \cite{ash01} for $\ g_{A}$=1.25.
The results of SSDH \cite{sim00} are in agreement with the experimental
half-lives of LBL \cite{gar93}, NEMO \cite{das95}, LBL collaboration \cite
{als97}, UC-Irvine \cite{sil97} and ITEP+INFN \cite{ash01}. The $%
T_{1/2}^{2\nu }$ calculated by Bobyk \textit{et al. }\cite{bob00} is in
agreement with all the experimental results due to a large range of values
from (5.04-16800)$\times $10$^{18}$ yr.

The 2$\nu $ $\beta \beta $ decay of $^{104}$Ru$\to ^{104}$Pd for $\ $0$%
^{+}\to $0$^{+}$ transition has not been experimentally investigated so far.
The theoretical calculations have been carried out in QRPA \cite{sta90} and
OEM \cite{hir94}. The predicted $T_{1/2}^{2\nu }$ in QRPA by Staudt \textit{%
et al.} \cite{sta90} is approximately one-fourth of our PHFB model
prediction for $g_{A}$=1.25 while the half life predicted by Hirsch \textit{%
et al.} in OEM \cite{hir94} is approximately 1.31 times larger. We predict a 
$T_{1/2}^{2\nu }$ for $^{104}$Ru to be 5.73$\times $10$^{22}$ yr for $g_{A}$%
=1.0.

The 2$\nu $ $\beta \beta $ decay of $^{110}$Pd$\to ^{110}$Cd for $\ $0$%
^{+}\to $0$^{+}$ transition has been investigated experimentally by Winter
only \cite{win52} long back and theoretically by employing QRPA \cite{sta90}%
, OEM \cite{hir94}, SRPA(WS) \cite{sto94} and SSDH \cite{civ98, sem00}. The $%
\beta \beta $ decay of $^{110}$Pd$\rightarrow ^{110}$Cd transition was
studied by Winter \cite{win52} deducing a half-life $T_{1/2}^{2\nu } > 6.0 \times $10$^{16}$ yr for 2$\nu $ $\beta \beta $ decay
mode and a total half-life $>$ 6.0$\times $10$^{17}$ yr for all
modes. The calculated $T_{1/2}^{2\nu }$ for $g_{A}=$1.25 in the present PHFB
model is 1.41$\times $10$^{20}$ yr, which is close to those of Semenov 
\textit{et al. }\cite{sem00} 1.6$\times $10$^{20}$ yr and twice of
Civitarese \textit{et al. }\cite{civ98} 0.7$\times $10$^{20}$ yr in SSDH. On
the other hand, the calculated half-life by Stoica \cite{sto94} in SRPA(WS)
is 1.186$\times $10$^{21}$ yr for the same $g_{A}$. The calculated average
half-life by Staudt \textit{et al. }\cite{sta90} in QRPA is 1.16$\times $10$%
^{19}$ yr and by Hirsch \textit{et al. }\cite{hir94} is 1.24$\times $10$%
^{21} $yr. For $g_{A}$=1.0, we predict a $T_{1/2}^{2\nu }$ for $^{110}$Pd to
be 3.44$\times $10$^{20}$ yr.

It is clear from the above discussions that the validity of nuclear models
presently employed to calculate the NTMEs $M_{2\nu }$ as well as half-lives $%
T_{1/2}^{2\nu }$ cannot be uniquely established due to large error bars in
experimental results as well as uncertainty in $\ g_{A}$. Further work is
necessary both in the experimental as well as theoretical front to judge the
relative applicability, success and failure of various nuclear models used
so far for the study of 2$\nu $ $\beta \beta $ decay processes.

\subsection{Deformation effect}

To understand the role of deformation on the NTME $M_{2\nu }$, we have
investigated the variation of the latter with respect to the change in
strength of the \textit{QQ} interaction $\chi _{qq}$. In Fig. 1, we have
displayed the dependence of $M_{2\nu }$ on the $\chi _{qq}$ for the 2$\nu $ $%
\beta \beta $ decay of $^{94,96}$Zr. In case of $^{94}$Zr, the $M_{2\nu }$
remains almost constant as the strength of $\chi _{qq}$ is changed from 0.00
to 0.80. As the strength of $\chi _{qq}$ is increased further up to 1.5, the 
$M_{2\nu }$ decreases except at 1.1, 1.3 and 1.5, where there is an increase
in the value of $M_{2\nu }.$ In case of $^{96}$Zr, the $M_{2\nu }$ remains
almost constant as the $\chi _{qq}$ is changed from 0.00 to 0.60. The $%
M_{2\nu }$ decreases as the $\chi _{qq}$ is changed to 1.20. As the $\chi
_{qq}$ is further varied to 1.5, the $M_{2\nu }$ increases initially and
remains almost constant. In case of $^{96}$Zr$\rightarrow ^{96}$Mo, the
experimental $M_{2\nu }$ is available. It is interesting to observe that the 
$M_{2\nu }$ gets tuned towards the realistic value as the $\chi _{qq}$
acquires a physical value around 1.0.

The dependence of $M_{2\nu }$ on the $\chi _{qq}$ has been displayed for the
2$\nu $ $\beta \beta $ decay of $^{98,100}$Mo in Fig. 2. In case of $^{98}$%
Mo, the $M_{2\nu }$ remains almost constant as the $\chi _{qq}$ is varied
from 0.00 to 0.60 and then decreases, while the $\chi _{qq}$ is changed to
1.2 except at 0.95. With further increase in $\chi _{qq}$, the $M_{2\nu }$
increases at $\chi _{qq}$=1.3 and 1.4 and then decreases at $\chi _{qq}$%
=1.5. In case of $^{100}$Mo, the $M_{2\nu }$ increases as the $\chi _{qq}$
is varied from 0.00 to 0.80 and then decreases, while the $\chi _{qq}$ is
changed to 1.10 except at 0.95. There is a further increase in $M_{2\nu }$
as the $\chi _{qq}$ is changed from 1.10 to 1.30 and then decreases up to
1.5. It is interesting to observe that in case of $^{100}$Mo$\rightarrow
^{100}$Ru, the $M_{2\nu }$ also gets tuned towards the realistic value as
the $\chi _{qq}$ acquires a physical value around 1.0.

In Fig. 3, we have displayed the dependence of $M_{2\nu }$ on the $\chi
_{qq} $ for the 2$\nu $ $\beta \beta $ decay of $^{104}$Ru and $^{110}$Pd.
The $M_{2\nu }$ remains almost constant as the $\chi _{qq}$ is varied from
0.00 to 0.60 and then decreases as the $\chi _{qq}$ is changed from 0.6 to
1.5 in case of $^{104}$Ru and $^{110}$Pd. To summarize, we have shown that
the deformations of the HFB intrinsic states play an important role in
reproducing a realistic $M_{2\nu }$.

To quantify the effect of deformation on $M_{2\nu }$, we define a quantity $%
D_{2\nu }$ as the ratio of $M_{2\nu }$ at zero deformation ($\chi _{qq}=0$)
and full deformation ($\chi _{qq}=1$). The $D_{2\nu }$ is given by 
\begin{equation}
D_{2\nu }=\frac{M_{2\nu }(\chi _{qq}=0)}{M_{2\nu }(\chi _{qq}=1)}
\end{equation}
The values of $D_{2\nu }$ are 2.29, 3.70, 1.86, 2.33, 5.47 and 3.14 for $%
^{94,96}$Zr,$^{98,100}$Mo,$^{104}$Ru and $\,^{110}$Pd nuclei respectively.
These values of $D_{2\nu }$ suggest that the $M_{2\nu }$ is quenched by a
factor of approximately 2 to 5.5 in the mass region 94$\leq A\leq $110 due
to deformation effects.

\section{CONCLUSIONS}

As a first step, we have tested the quality of HFB wave functions by
comparing the theoretically calculated results for a number of spectroscopic
properties of $^{94,96}$Zr, $^{94,96,98,100}$Mo, $^{98,100,104}$Ru, $%
^{104,110}$Pd and $^{110}$Cd\ nuclei with the available experimental data.
To be more specific, we have computed the yrast spectra, reduced $B(E2$:$%
0^{+}\rightarrow 2^{+})$ transition probabilities, quadrupole moments $%
Q(2^{+})$ and $g$-factors $g(2^{+})$. Subsequently, 
the
reliability of the
intrinsic wave functions has been tested by calculating the $M_{2\nu }$ of $%
^{96}$Zr and $^{100}$Mo, for which the 2$\nu $ $\beta \beta $\ decay 
has already been measured. 
In case of $^{96}$Zr and $^{100}$Mo, the
agreement between the theoretically calculated and experimentally observed $%
M_{2\nu }$ as well as $T_{1/2}^{2\nu }$ makes us confident to predict the
half-lives $T_{1/2}^{2\nu }$ for other nuclei undergoing 2$\nu $ $\beta
\beta $\ decay in the mass region 94$\leq A\leq $110. For $^{94}$Zr,$^{98}$%
Mo,$^{104}$Ru and $\,^{110}$Pd isotopes, the values of $T_{1/2}^{2\nu }$ for 
$g_{A}$=1.25-1.00 are (7.51-18.34)$\times $10$^{22}$ yr, (6.09-14.87)$\times 
$10$^{29}$ yr, (2.35-5.73)$\times $10$^{22}$ yr and (1.41-3.44)$\times $10$%
^{20}$ yr respectively.

Further, we have shown that the deformations of the intrinsic ground states
of $^{96}$Zr, $^{96,100}$Mo and $^{100}$Ru play a crucial role in
reproducing a realistic NTME in case of $^{96}$Zr and $^{100}$Mo. The NTMEs $%
M_{2\nu }$ are quenched by a factor of approximately 2 to 5.5 in the mass
region 94$\leq A\leq $110 due to the deformation. A reasonable agreement
between the calculated and observed spectroscopic properties of $^{94,96}$%
Zr, $^{94,96,98,100}$Mo, $^{98,100,104}$Ru, $^{104,110}$Pd and $^{110}$Cd as
well as the $\ $2$\nu $ $\beta \beta $\ decay rate of $^{94,96}$Zr,$%
^{98,100} $Mo,$^{104}$Ru and $\,^{110}$Pd makes us confident to employ the
same PHFB wave functions to study the $\ $0$\nu $ $\beta \beta $\ decay,
which will be communicated in the future.

\textbf{Acknowledgment:} P. K. Rath will like to acknowledge the financial
support provided by CTS, IIT, Kharagpur, where the present work has been
finalized. Further, R. Chandra is grateful to CSIR, India for providing
Senior research fellowship vide award no. 9/107(222)/2KI/EMR-I.

\noindent \textbf{Table I.} Excitation energies (in MeV) of $J^{\pi }=2^{+},$
$4^{+},$ $6^{+}$ yrast states of some nuclei in the mass range 94$\leq A\leq 
$ 110 with fixed $Gp$= 30/\textit{A}, $Gn$= 20/\textit{A }(22/\textit{A} for 
$^{96}$Zr) and $\varepsilon $(\textit{h}$_{11/2}$)=8.6 MeV.

\begin{tabular}{llllllllll}
\hline\hline
Nucleus & $\chi _{pn}$ & \multicolumn{2}{c}{Theory} & Experiment \cite{sak84}
& Nucleus & $\chi _{pn}$ & \multicolumn{2}{c}{Theory} & Experiment \cite
{sak84} \\ \hline
$^{94}$Zr & 0.02519 & $E_{2^{+}}$ & 0.9182 & 0.9183 & $^{94}$Mo & 0.02670 & $%
E_{2^{+}}$ & 0.8715 & 0.871099 \\ 
&  & $E_{4^{+}}$ & 1.9732 & 1.4688 &  &  & $E_{4^{+}}$ & 1.9685 & 1.573726
\\ 
&  & $E_{6^{+}}$ & 2.7993 &  &  &  & $E_{6^{+}}$ & 3.3136 & 2.42337 \\ \hline
$^{96}$Zr & 0.01717 & $E_{2^{+}}$ & 1.7570 & 1.7507 & $^{96}$Mo & 0.02557 & $%
E_{2^{+}}$ & 0.7779 & 0.778213 \\ 
&  & $E_{4^{+}}$ & 3.5269 & 3.1202 &  &  & $E_{4^{+}}$ & 2.0373 & 1.62815 \\ 
&  & $E_{6^{+}}$ & 9.7261 &  &  &  & $E_{6^{+}}$ & 3.5776 & 2.44064 \\ \hline
$^{98}$Mo & 0.01955 & $E_{2^{+}}$ & 0.7892 & 0.78742 & $^{98}$Ru & 0.02763 & 
$E_{2^{+}}$ & 0.6513 & 0.65241 \\ 
&  & $E_{4^{+}}$ & 1.9522 & 1.51013 &  &  & $E_{4^{+}}$ & 1.9430 & 1.3978 \\ 
&  & $E_{6^{+}}$ & 3.3098 & 2.3438 &  &  & $E_{6^{+}}$ & 3.6548 & 2.2227 \\ 
\hline
$^{100}$Mo & 0.01906 & $E_{2^{+}}$ & 0.5356 & 0.53555 & $^{100}$Ru & 0.01838
& $E_{2^{+}}$ & 0.5395 & 0.53959 \\ 
&  & $E_{4^{+}}$ & 1.4719 & 1.13594 &  &  & $E_{4^{+}}$ & 1.5591 & 1.2265 \\ 
&  & $E_{6^{+}}$ & 2.6738 &  &  &  & $E_{6^{+}}$ & 2.8940 & 2.0777 \\ \hline
$^{104}$Ru & 0.02110 & $E_{2^{+}}$ & 0.3580 & 0.35799 & $^{104}$Pd & 0.01486
& $E_{2^{+}}$ & 0.5552 & 0.55579 \\ 
&  & $E_{4^{+}}$ & 1.1339 & 0.8885 &  &  & $E_{4^{+}}$ & 1.5729 & 1.32359 \\ 
&  & $E_{6^{+}}$ & 2.2280 & 1.5563 &  &  & $E_{6^{+}}$ & 2.8790 & 2.2498 \\ 
\hline
$^{110}$Pd & 0.01417 & $E_{2^{+}}$ & 0.3737 & 0.3738 & $^{110}$Cd & 0.01412
& $E_{2^{+}}$ & 0.6576 & 0.657751 \\ 
&  & $E_{4^{+}}$ & 1.1563 & 0.9208 &  &  & $E_{4^{+}}$ & 1.8709 & 1.542412
\\ 
&  & $E_{6^{+}}$ & 2.2254 & 1.5739 &  &  & $E_{6^{+}}$ & 3.3865 & 2.479893
\\ \hline\hline
\end{tabular}

\pagebreak 

\begin{tabular}{c}
\end{tabular}

\noindent \textbf{Table II.} Comparison of calculated and experimentally
observed reduced transition probabilities $B(E2$:$0^{+}\rightarrow 2^{+})$
in $e^{2}$ b$^{2}$, static quadrupole moments \ $Q(2^{+})$ in $e$ b and $g$%
-factors $g(2^{+})$ in nuclear magneton. Here $B(E2)$ and $Q(2^{+})$ are
calculated for effective charge $e_{p}=$1+$e_{eff}$ and $e_{n}=e_{eff}$. The 
$g(2^{+})$ has been calculated for $g_{l}^{\pi }=$1.0, $g_{l}^{\nu }=$0.0
and $g_{s}^{\pi }=g_{s}^{\nu }=$0.60.

\begin{tabular}{cllllllllll}
\hline\hline
Nucleus & \multicolumn{4}{c}{$B(E2$:$0^{+}\rightarrow 2^{+})$ \ \ \ \ \ \ \
\ \ } & \multicolumn{4}{c}{$Q(2^{+})$\ \ \ \ \ \ \ \ \ \ \ \ \ \ \ } & 
\multicolumn{2}{c}{$g(2^{+})\ $\ \ \ \ \ } \\ 
\multicolumn{1}{l}{} & \multicolumn{3}{l}{\ \ \ \ \ \ \ \ \ \ \ \ Theory} & 
\ \ \ Experiment$^{a}$ & \multicolumn{3}{l}{\ \ \ \ \ \ \ \ \ \ Theory} & \
Experiment$^{b}$ & Theory & Experiment$^{b}$ \\ 
& \multicolumn{3}{c}{$e_{eff}$} &  & \multicolumn{3}{c}{$e_{eff}$} &  &  & 
\\ 
\multicolumn{1}{l}{} & 0.40 & 0.50 & 0.60 &  & 0.40 & 0.50 & 0.60 &  &  & 
\\ \hline
\multicolumn{1}{l}{$^{94}$Zr} & 0.046 & 0.062 & \textbf{0.081} & 0.081$\pm $%
0.017 & -0.168 & -0.195 & \textbf{-0.222} &  & 0.121 & -0.329$\pm $0.015$^{c}
$ \\ 
\multicolumn{1}{l}{} &  &  &  & 0.066$\pm $0.014 &  &  &  &  &  & -0.26$\pm $%
0.06 \\ 
\multicolumn{1}{l}{} &  &  &  & 0.056$\pm $0.014 &  &  &  &  &  & -0.05$\pm $%
0.05 \\ 
\multicolumn{1}{l}{$^{94}$Mo} & 0.148 & 0.188 & \textbf{0.232} & 0.230$\pm $%
0.040 & -0.347 & -0.391 & \textbf{-0.435} & -0.13$\pm $0.08 & 0.343 &  \\ 
\multicolumn{1}{l}{} &  &  &  & 0.270$\pm $0.035 &  &  &  &  &  &  \\ 
\multicolumn{1}{l}{} &  &  &  & 0.290$\pm $0.044 &  &  &  &  &  &  \\ \hline
\multicolumn{1}{l}{$^{96}$Zr} & 0.044 & \textbf{0.060} & 0.078 & 0.055$\pm $%
0.022 & -0.012 & \textbf{-0.015} & -0.018 &  & 0.254 &  \\ 
\multicolumn{1}{l}{$^{96}$Mo} & 0.265 & \textbf{0.335} & 0.413 & 0.310$\pm $%
0.047 & -0.466 & \textbf{-0.524} & -0.582 & -0.20$\pm $0.08 & 0.563 &  \\ 
\multicolumn{1}{l}{} &  &  &  & 0.302$\pm $0.039 &  &  &  &  &  &  \\ 
\multicolumn{1}{l}{} &  &  &  & 0.288$\pm $0.016 &  &  &  &  &  &  \\ \hline
\multicolumn{1}{l}{$^{98}$Mo} & \textbf{0.234} & 0.302 & 0.378 & 0.260$\pm $%
0.040 & \textbf{-0.439} & -0.498 & -0.557 & -0.26$\pm $0.09 & 0.376 & 0.34$%
\pm $0.18 \\ 
\multicolumn{1}{l}{} &  &  &  & 0.270$\pm $0.040 &  &  &  &  &  &  \\ 
\multicolumn{1}{l}{} &  &  &  & 0.267$\pm $0.005 &  &  &  &  &  &  \\ 
\multicolumn{1}{l}{$^{98}$Ru} & \textbf{0.433} & 0.543 & 0.665 & 0.411$\pm $%
0.035 & \textbf{-0.596} & -0.667 & -0.739 & -0.20$\pm $0.09 & 0.528 & 0.39$%
\pm $0.30 \\ 
\multicolumn{1}{l}{} &  &  &  & 0.475$\pm $0.038 &  &  &  & -0.03$\pm $0.14
&  &  \\ 
\multicolumn{1}{l}{} &  &  &  & 0.392$\pm $0.012 &  &  &  &  &  &  \\ 
\hline\hline
\end{tabular}

\smallskip continued.....

\pagebreak 

\smallskip Table II continued....

\begin{tabular}{lllllllllll}
\hline
$^{100}$Mo & 0.320 & 0.412 & \textbf{0.515} & 0.511$\pm $0.009 & -0.512 & 
-0.581 & \textbf{-0.650} & -0.42$\pm $0.09 & 0.477 & 0.34$\pm $0.18 \\ 
&  &  &  & 0.516$\pm $0.010 &  &  &  & -0.39$\pm $0.08 &  &  \\ 
&  &  &  & 0.470$\pm $0.024 &  &  &  &  &  &  \\ 
$^{100}$Ru & 0.308 & 0.393 & \textbf{0.488} & 0.494$\pm $0.006 & -0.503 & 
-0.568 & \textbf{-0.633} & -0.54$\pm $0.07 & 0.355 & 0.42$\pm $0.03 \\ 
&  &  &  & 0.493$\pm $0.003 &  &  &  & -0.40$\pm $0.12 &  & 0.47$\pm $0.06
\\ 
&  &  &  & 0.501$\pm $0.010 &  &  &  & -0.43$\pm $0.07 &  &  \\ \hline
$^{104}$Ru & 0.572 & 0.732 & \textbf{0.912} & 0.93$\pm $0.06 & -0.684 & 
-0.774 & \textbf{-0.864} & -0.76$\pm $0.19 & 0.339 & 0.41$\pm $0.05 \\ 
&  &  &  & 1.04$\pm $0.16 &  &  &  & -0.70$\pm $0.08 &  &  \\ 
&  &  &  & 0.841$\pm $0.016 &  &  &  & -0.66$\pm $0.05 &  &  \\ 
$^{104}$Pd & 0.361 & 0.460 & \textbf{0.571} & 0.547$\pm $0.038 & -0.543 & 
-0.613 & \textbf{-0.682} & -0.47$\pm $0.10 & 0.439 & 0.46$\pm $0.04 \\ 
&  &  &  & 0.61$\pm $0.09 &  &  &  &  &  & 0.40$\pm $0.05 \\ 
&  &  &  & 0.535$\pm $0.035 &  &  &  &  &  & 0.38$\pm $0.04 \\ \hline
$^{110}$Pd & 0.479 & \textbf{0.614} & 0.766 & 0.780$\pm $0.120 & -0.626 & 
\textbf{-0.708} & -0.791 & -0.72$\pm $0.14 & 0.478 & 0.37$\pm $0.03 \\ 
&  &  &  & 0.820$\pm $0.080 &  &  &  & -0.55$\pm $0.08 &  & 0.35$\pm $0.03
\\ 
&  &  &  & 0.860$\pm $0.060 &  &  &  & -0.47$\pm $0.03 &  & 0.31$\pm $0.03
\\ 
$^{110}$Cd & 0.427 & \textbf{0.548} & 0.685 & 0.504$\pm $0.040 & -0.590 & 
\textbf{-0.668} & -0.746 & -0.40$\pm $0.04 & 0.358 & 0.31$\pm $0.07 \\ 
&  &  &  & 0.467$\pm $0.019 &  &  &  & -0.39$\pm $0.06 &  & 0.28$\pm $0.05
\\ 
&  &  &  & 0.450$\pm $0.020 &  &  &  & -0.36$\pm $0.08 &  & 0.285$\pm $0.055
\\ \hline\hline
\end{tabular}

$^{a}$Reference \cite{ram87}; $^{b}$Reference \cite{rag89}; $^{c}$Reference 
\cite{spe02}

\pagebreak 

\noindent \textbf{Table III. }Experimentally observed and theoretically
calculated $M_{2\nu }$ and half-lives $T_{1/2}^{2\nu }$ for 0$%
^{+}\rightarrow $0$^{+}$ transition of $^{94,96}$Zr, $^{98,100}$Mo,$^{104}$%
Ru and $^{110}$Pd nuclei in different nuclear models. The numbers
corresponding to (a) and (b) are calculated for $g_{A}=$1.25 and 1.0
respectively.

\begin{tabular}{llllllcllll}
\hline\hline
Nuclei &  & \multicolumn{4}{c}{Experiment} &  & \multicolumn{4}{c}{Theory}
\\ 
& Ref. & Projects & $T_{1/2}^{2\nu }$(yr) &  & $\left| M_{2\nu }\right| $ & 
Ref. & Models & $\left| M_{2\nu }\right| $ &  & $T_{1/2}^{2\nu }$(yr) \\ 
\hline
$^{94}$Zr & \cite{arn99} & NEMO & $>$ 1.1$\times $10$^{-5}$ & (a) & $<$62.815
& * & PHFB & 0.076 & (a) & 7.51 \\ 
(10$^{22}$ yr) &  &  &  & (b) & $<$ 98.148 &  &  &  & (b) & 18.34 \\ 
&  &  &  &  &  & \cite{bob00} & SRQRPA &  &  & 3.08-659 \\ 
&  &  &  &  &  & \cite{hir94} & OEM &  &  & 168 \\ 
&  &  &  &  &  & \cite{sta90} & QRPA &  &  & 6.93 \\ \hline
&  &  &  &  &  &  &  &  &  &  \\ 
$^{96}$Zr & \cite{wie01} & $^{\dagger }$gch. & 0.94$\pm $0.32 & (a) & 0.074$%
_{-0.010}^{+0.017}$ & * & PHFB & 0.058 & (a) & 1.56 \\ 
(10$^{19}$ yr) &  &  &  & (b) & 0.116$_{-0.016}^{+0.027}$ &  &  &  & (b) & 
3.80 \\ 
& \cite{arn99} & NEMO & 2.1$_{-0.4}^{+0.8}\pm $0.2 & (a) & 0.050$%
_{-0.009}^{+0.009}$ & \cite{bob00} & SRQRPA &  &  & 0.452-61 \\ 
&  &  &  & (b) & 0.078$_{-0.014}^{+0.014}$ & \cite{rum98} & SU(4)$_{\sigma
\tau }$ & 0.0678 & (a) & 1.13 \\ 
& \cite{bar98} & NEMO & 2.0$_{-0.5}^{+0.9}\pm $0.5 & (a) & 0.051$%
_{-0.012}^{+0.021}$ &  &  &  & (b) & 2.76 \\ 
&  &  &  & (b) & 0.080$_{-0.019}^{+0.033}$ & \cite{toi97} & RQRPA(WS) &  & 
& 4.2 \\ 
& \cite{kaw93} & $^{\dagger }$gch. & 3.9$\pm $0.9 & (a) & 0.036$%
_{-0.004}^{+0.005}$ & \cite{toi97} & RQRPA(AWS) &  &  & 4.4 \\ 
&  &  &  & (b) & 0.057$_{-0.006}^{+0.008}$ & \cite{bar96} & QRPA(AWS) & 
0.12-0.31 & (a) & 0.054-0.36 \\ 
& \cite{ell02} & Average & 1.4$_{-0.5}^{+3.5}$ & (a) & 0.061$%
_{-0.028}^{+0.015}$ &  &  &  & (b) & 0.13-0.88 \\ 
&  & Value &  & (b) & 0.095$_{-0.044}^{+0.024}$ & \cite{sto95} & SRPA(WS) & 
0.022 & (a) & 10.72 \\ 
& \cite{bar02} & Recommended & 2.1$_{-0.4}^{+0.8}$ & (a) & 0.050$%
_{-0.007}^{+0.006}$ &  &  &  & (b) & 26.18 \\ 
&  & Value &  & (b) & 0.078$_{-0.012}^{+0.009}$ & \cite{hir94} & OEM &  &  & 
20.2 \\ 
&  &  &  &  &  & \cite{sta90} & QRPA &  &  & 1.08 \\ 
&  &  &  &  &  & \cite{eng88} & QRPA & 0.124 & (a) & 0.34 \\ 
&  &  &  &  &  &  &  &  & (b) & 0.82 \\ 
\hline
\end{tabular}
continued..\pagebreak

table III continued.

\begin{tabular}{llllllcllll}
\hline
$^{98}$Mo &  &  &  &  &  & * & PHFB & 0.130 & (a) & 6.09 \\ 
(10$^{29}$ yr) &  &  &  &  &  &  &  &  & (b) & 14.87 \\ 
&  &  &  &  &  & \cite{bob00} & SRQRPA &  &  & 4.06-15.2 \\ 
&  &  &  &  &  & \cite{hir94} & OEM &  &  & 61.6 \\ 
&  &  &  &  &  & \cite{sta90} & QRPA &  &  & 29.6 \\
 \hline

$^{100}$Mo & \cite{ash01} & ITEP+INFN & 7.2$\pm $0.9$\pm $1.8 & (a) & 0.121$%
_{-0.018}^{+0.032}$ & * & PHFB & 0.104 & (a) & 9.79 \\ 
(10$^{18}$ yr) &  &  &  & (b) & 0.190$_{-0.028}^{+0.050}$ &  &  &  & (b) & 
23.90 \\ 
& \cite{ash99} & ITEP & 8.5 & (a) & 0.112 & \cite{sim00} & SSDH &  & (a) & 
7.15-8.97 \\ 
&  &  &  & (b) & 0.174 & \cite{bob00} & SRQRPA &  &  & 5.04-16800 \\ 
& \cite{sil97} & UC Irvine & 6.82$_{-0.53}^{+0.38}\pm $0.68 & (a) & 0.125$%
_{-0.009}^{+0.013}$ & \cite{rum98} & SU(4)$_{\sigma \tau }$ & 0.1606 & (a) & 
4.11 \\ 
&  &  &  & (b) & 0.195$_{-0.014}^{+0.020}$ &  &  &  & (b) & 10.03 \\ 
& \cite{als97} & LBL+MHC+ & 7.6$_{-1.4}^{+2.2}$ & (a) & 0.118$%
_{-0.014}^{+0.013}$ & \cite{civ98} & SSDH & 0.18 & (a) & 3.27 \\ 
&  & UNM+INEL &  & (b) & 0.185$_{-0.022}^{+0.020}$ &  &  &  & (b) & 7.99 \\ 
& \cite{das95} & NEMO & 9.5$\pm $0.4$\pm $0.9 & (a) & 0.106$%
_{-0.007}^{+0.008}$ & \cite{sto95} & SRPA(WS) & 0.059 & (a) & 30.45 \\ 
&  &  &  & (b) & 0.165$_{-0.010}^{+0.013}$ &  &  &  & (b) & 74.34 \\ 
& \cite{gar93} & LBL & 9.7$\pm $4.9 & (a) & 0.105$_{-0.019}^{+0.044}$ & \cite
{hir95b} & SU(3)(SPH) & 0.152 & (a) & 4.59 \\ 
&  &  &  & (b) & 0.163$_{-0.030}^{+0.069}$ &  &  &  & (b) & 11.2 \\ 
& \cite{eji91} & ELEGANTS V & 11.5$_{-2.0}^{+3.0}$ & (a) & 0.096$%
_{-0.011}^{+0.010}$ & \cite{hir95b} & SU(3)(DEF) & 0.108 & (a) & 9.09 \\ 
&  &  &  & (b) & 0.150$_{-0.016}^{+0.015}$ &  &  &  & (b) & 22.19 \\ 
& \cite{ell91} & UC Irvine & 11.6$_{-0.8}^{+3.4}$ & (a) & 0.096$%
_{-0.012}^{+0.004}$ & \cite{hir94} & OEM &  &  & 35.8 \\ 
&  &  &  & (b) & 0.149$_{-0.018}^{+0.005}$ & \cite{suh94} & QRPA(EMP) & 0.101
& (a) & 10.39 \\ 
& \cite{vas90} & INS Baksan & 3.3$_{-1.0}^{+2.0}$ & (a) & 0.179$%
_{-0.038}^{+0.036}$ &  &  &  & (b) & 25.37 \\ 
&  &  &  & (b) & 0.280$_{-0.059}^{+0.055}$ & \cite{gri92} & QRPA(EMP) & 0.256
& (a) & 1.62 \\ 
& \cite{ell02} & Average & 8.0$\pm $0.6 & (a) & 0.115$_{-0.004}^{+0.005}$ & 
&  &  & (b) & 3.95 \\ 
&  & Value &  & (b) & 0.180$_{-0.006}^{+0.007}$ & \cite{sta90} & QRPA &  & 
& 1.13 \\ 
& \cite{bar02} & Average & 8.0$\pm $0.7 & (a) & 0.115$_{-0.005}^{+0.005}$ & 
\cite{eng88} & QRPA & 0.211 & (a) & 2.38 \\ 
&  & Value &  & (b) & 0.180$_{-0.007}^{+0.008}$ &  &  &  & (b) & 5.81 \\ 
\hline
\end{tabular}

\smallskip ...continued\pagebreak 

\smallskip Table III continued..

\begin{tabular}{llllllcllll}
\hline
&  &  &  &  &  &  &  &  &  &  \\ 
$^{104}$Ru &  &  &  &  &  & * & PHFB & 0.068 & (a) & 2.35 \\ 
(10$^{22}$ yr) &  &  &  &  &  &  &  &  & (b) & 5.73 \\ 
&  &  &  &  &  & \cite{hir94} & OEM &  &  & 3.09 \\ 
&  &  &  &  &  & \cite{sta90} & QRPA &  &  & 0.629 \\ \hline
&  &  &  &  &  &  &  &  &  &  \\ 
$^{110}$Pd & \cite{win52} &  & $>$ 6.0$\times $10$^{-4}$ & (a) & $<$ 6.468 & 
* & PHFB & 0.133 & (a) & 1.41 \\ 
(10$^{20}$ yr) &  &  &  & (b) & $<$ 10.106 &  &  &  & (b) & 3.44 \\ 
&  &  &  &  &  & \cite{sem00} & SSDH &  &  & 1.6 \\ 
&  &  &  &  &  & \cite{civ98} & SSDH & 0.19 & (a) & 0.7 \\ 
&  &  &  &  &  &  &  &  & (b) & 1.70 \\ 
&  &  &  &  &  & \cite{sto94} & SRPA(WS) & 0.046 & (a) & 11.86 \\ 
&  &  &  &  &  &  &  &  & (b) & 28.96 \\ 
&  &  &  &  &  & \cite{hir94} & OEM &  &  & 12.4 \\ 
&  &  &  &  &  & \cite{sta90} & QRPA &  &  & 0.116 \\ \hline\hline
\end{tabular}

$^{\dagger }$gch. denotes geochemical experiment and $^{*}$Present work.

\end{document}